\documentclass{iaus}
\usepackage{graphicx}

\title[short title of paper] 
{First astronomical detection of the CF$^+$ ion}

\author[Neufeld et al.]   
{D.~A.~Neufeld$^1$, P.~Schilke$^2$, K.~M.~Menten$^2$, M.~G.~Wolfire$^3$, 
\break J.~H.~Black$^4$, F.~Schuller$^2$,  H.~M\"uller$^{2,5}$, S.~Thorwirth$^2$,
R.~G\"usten$^2$, S.~Philipp$^2$}

\affiliation{$^1$Department of Physics and Astronomy, Johns Hopkins University, Baltimore, USA\\[\affilskip]
$^2$ Max-Planck-Institut f\"ur Radioastronomie, Bonn, Germany\\[\affilskip]
$^3$ Astronomy Department, University of Maryland, College Park, USA\\[\affilskip]
$^4$ Onsala Space Observatory, Onsala, Sweden \\[\affilskip]
$^5$ Universit\"at zu K\"oln, K\"oln, Germany \\[\affilskip]}

\pubyear{2005}
\volume{231}  
\date{21 September 2005}
\jname{Astrochemistry: Recent Successes and Current Challenges}
\editors{D.~C.~Lis, G.~A.~Blake, and E.~Herbst}
\begin{document}

\maketitle

\begin{abstract}
We report the first astronomical detection of the CF$^+$ (fluoromethylidynium) ion,
obtained by recent observations of its $J = 1 - 0$ (102.6~GHz), $J = 2 - 1$
(205.2~GHz) , and $J = 3 - 2$ (307.7~GHz) pure rotational emissions toward the
Orion Bar.  Our search for CF$^+$ -- carried out using the IRAM 30m and APEX 12m
telescopes -- was motivated by recent theoretical models that predict CF$^+$
abundances of a $\rm few \times 10^{-10}$ in UV-irradiated molecular regions where C$^+$ is
present.  The measurements confirm the predictions.  They provide support for our current theories of
interstellar fluorine chemistry, which suggest that hydrogen fluoride should
be ubiquitous in interstellar gas clouds.

\end{abstract}

The fluoromethylidynium ion, CF$^+$, sounds like a rather exotic species, but in some ways 
it is very familiar, being isoelectronic with carbon monoxide, the most widely 
observed interstellar molecule.  Indeed, adding a proton and 
two neutrons to the oxygen nucleus in CO would yield CF$^+$.  The rotational constant is smaller by about 
10\%, and the dipole moment larger by a factor of 10, but there are still 14 electrons 
in a $^1\Sigma^+$ ground state.  Furthermore, the $J=1-0$, $J=2-1$ and $J=3-2$
rotational transitions still lie respectively in the 3, 1.5, and 1~mm bands
accessible to ground-based observatories.

\begin{figure}
\includegraphics[height=4.7in,width=3.1in,angle=270]{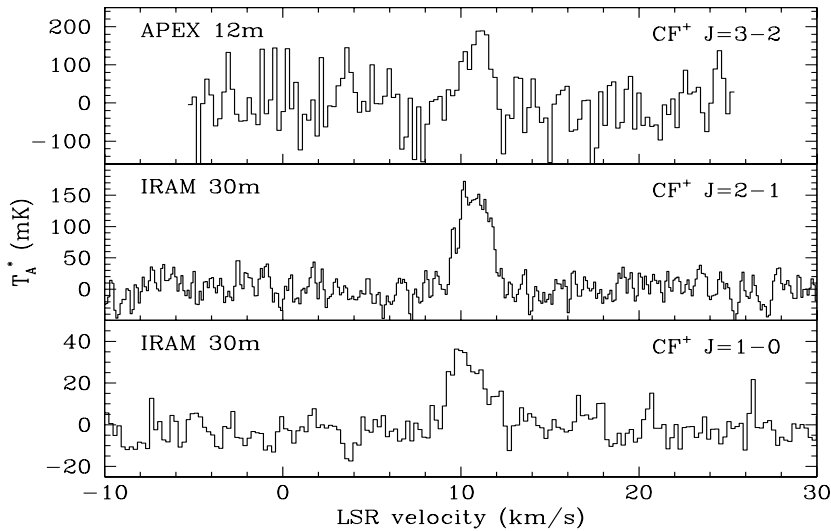}
  \caption{CF$^+$ $J=1-0$, $J=2-1$, and $J=3-2$ spectra obtained
toward the Orion Bar}
\end{figure}

Figure 1 shows the $J=1-0$, $J=2-1$, and $J=3-2$ spectra obtained
toward one of several positions where CF$^+$ has been detected in the Orion Bar $\rm (\alpha=05h\,35m\,22.80s, \delta= -05d\,25^\prime\,01.0^{\prime\prime})$, 
a well-studied photodissociation region with an edge-on geometry that is 
favorable for the detection of molecules of low abundance.  The lower two transitions were observed
at the IRAM 30~m telescope, while the submillimeter $J=3-2$ line was detected at
the new APEX 12~m telescope, located at an altitude of 5100 m on the Chajnantor plateau in the Atacama desert.

When we construct a rotational diagram, we find -- rather fortunately -- that the $J=1$, $J=2$, and $J=3$ 
states that we have observed are almost certainly the three most highly-populated rotational states, and 
that for any reasonable extrapolation to higher $J$, the total 
CF$^+$ column density lies close to 
$1.7 \times 10^{12}\rm \,cm^{-2}$, averaged over our beam.

Our discovery of CF$^+$ was not obtained by serendipity, nor in a line survey, but rather through a targeted search that was prompted by a recent theoretical study undertaken by
Neufeld, Wolfire \& Schilke (2005).  Motivated by the detection
of HF toward the Sgr B2 cloud (Neufeld et al.\ 1997), we considered the chemistry 
of interstellar fluorine-bearing molecules and reached three key conclusions:

(1) Hydrogen fluoride forms rapidly by the reaction of fluorine atoms with H$_2$.  
This is a simple consequence of thermochemistry: HF has the highest 
dissociation energy of any neutral diatomic hydride, and is the only such molecule to be 
more strongly bound than molecular hydrogen.  Fluorine is therefore the only element in the periodic table to 
have a neutral atom that reacts exothermically with H$_2$.  As Zhu et al.\ (2002) have 
shown, the reaction is expected to be fairly rapid even in cold molecular clouds.

(2) Hydrogen fluoride is
the dominant reservoir of fluorine nuclei (solar abundance: $n_F/n_H \sim 3 \times 10^{-8}$) in the gas phase, even near cloud 
surfaces.  Beneath the surface of a UV-illuminated cloud, HF forms 
precisely where hydrogen becomes molecular, and long before carbon gets 
incorporated into CO.  

(3) Hydrogen fluoride is destroyed primarily by 
photodissociation and by reaction with C$^+$ to form CF$^+$.  There is a 
substantial region where C$^+$ and HF overlap, and it is in this region where the CF$^+$ 
abundance is largest, accounting for a few percent of the total fluorine nuclei.  
The total predicted CF$^+$  
column density is $\sim 10^{12} \rm \,cm^{-2}$ over a wide range of physical conditions.

To summarize, we have obtained the first astronomical detection of the CF$^+$ ion, toward the Orion Bar.  This observation supports a theoretical model that predicts large abundances of HF close to cloud surfaces where the C$^+$ abundance is high, and it suggests that Herschel and SOFIA will detect widespread absorption by the $J=1-0$ transition of HF in diffuse clouds along lines-of-sight to 
far-infrared continuum sources. 

\begin{acknowledgments}
IRAM is supported by INSU/CNRS (France), MPG (Germany), and IGN
(Spain). APEX is a joint project of the Max Planck Institute for Radio
Astronomy, the European Southern Observatory, and the
Swedish National Facility for Radio Astronomy.
\end{acknowledgments}

\end{document}